\documentclass[3p,times,procedia]{elsarticle}
\usepackage{nupha_ecrc}

\volume{00}

\firstpage{1}

\journalname{Nuclear Physics A}

\runauth{J.~Noronha and G.~S.~Denicol}

\jid{nupha}

\jnltitlelogo{Nuclear Physics A}

\usepackage{amssymb}
\usepackage{amsmath}

\usepackage[figuresright]{rotating}

\begin{document}

\begin{frontmatter}



\dochead{XXVIth International Conference on Ultrarelativistic Nucleus-Nucleus Collisions\\ (Quark Matter 2017)}

\title{The onset of fluid-dynamical behavior in relativistic kinetic
theory}


\author[a]{Jorge Noronha}
\address[a]{Instituto de F\'{i}sica, Universidade de S\~{a}o Paulo, USP, 05315-970 S\~{a}o Paulo, SP, Brazil}

\author[b]{Gabriel S.~Denicol}
\address[b]{Instituto de F\'isica, Universidade Federal Fluminense,
UFF, Niter\'oi, 24210-346, RJ, Brazil}

\begin{abstract}
In this proceedings we discuss recent findings regarding the large order behavior of the Chapman-Enskog expansion in relativistic kinetic theory. It is shown that this series in powers of the Knudsen number has zero radius of convergence in the case of a Bjorken expanding fluid described by the Boltzmann equation in the relaxation time approximation. This divergence stems from the presence of non-hydrodynamic modes, which give non-perturbative contributions to the Knudsen series.    

\end{abstract}


\end{frontmatter}


\section{Introduction}
\label{intro}

Relativistic hydrodynamics is the main tool used in the description of the spacetime evolution of the quark-gluon plasma formed in ultrarelativistic heavy ion collisions \cite{Heinz:2013th}. Following the seminal work by Hilbert \cite{hilbert}, Chapman and Enskog \cite{chapman-cowling} in the non-relativistic limit, hydrodynamics has since then been understood \cite{landau} as an effective theory whose accuracy is controlled by the Knudsen number $K_N \sim \lambda/L$, a ratio between a microscopic scale $\lambda$ (e.g., the mean free path in gases) and a macroscopic scale that characterizes the spatial gradients of hydrodynamic fields such as the temperature $T$ and flow velocity $u_\mu$. Fluid dynamics is expected to provide a good description of the system's evolution when there is a large separation of scales such that $K_N \ll 1$. In this case, one expects that the equations of hydrodynamics may be organized order by order in powers of $K_N$: at 0th-order one finds ideal fluid dynamics while at 1st order in $K_N$ one obtains the relativistic version of Navier-Stokes equations \cite{landau}. In principle, one may continue this expansion (though in practice this is very hard to pursue in the absence of powerful symmetry constraints such as Weyl invariance \cite{Baier:2007ix}), with the hope that it provides a better description of the fluid.  

The large spatial gradients expected to occur in the early stages of the quark-gluon plasma formed in heavy ion collisions \cite{Schenke:2012wb,Niemi:2014wta,Noronha-Hostler:2015coa}, and the collective behavior present in small collision systems such as pA \cite{Schenke:2017bog} where such a large separation of scales may not take place, motivate one to understand how relativistic hydrodynamic behavior emerges as an effective description of rapidly evolving kinetic systems \cite{Denicol:2014xca,Denicol:2014tha}. In this proceedings we tackle this question using a simple ``toy model" of the quark-gluon plasma defined within kinetic theory and study the large order behavior of the Chapman-Enskog gradient series. 

\section{Kinetic model}
\label{kineticmodel}

We consider a longitudinally expanding system of massless particles
undergoing Bjorken flow \cite{bjorken} in Minkowski spacetime described using the coordinates $x^\mu = (\tau, x,y,\eta)$ with $\tau=\sqrt{t^2-z^2}>0$ and $\eta = \tanh^{-1}(z/t)$. In this case all the quantities depend only on $\tau$ and, even though in these coordinates the system is homogeneous and $u^{\mu }=\left( 1,0,0,0\right)$, the fluid experiences a nonzero expansion rate $\theta \equiv \partial_\mu u^\mu= 1/\tau$. We employ the Boltzmann equation in the relaxation time approximation (RTA) \cite{Denicol:2014xca,Denicol:2014tha,AW}
\begin{equation}
\partial _{\tau }f_{\mathbf{k}}=- \frac{1}{\tau_R}\, \left(f_\mathbf{k}-f_{\mathrm{eq}}\right),
\label{eqBolt}
\end{equation}%
where $f_{\mathbf{k}}\equiv f\left(\tau,k_\eta,k_0\right)$ is the single particle distribution function, the 4-momentum is $k_{\mu }=(k_{0},\mathbf{k})$ and $k_{0}=\sqrt{k_{x}^{2}+k_{y}^{2}+k_{\eta
}^{2}/\tau ^{2}}$, $\tau_R$ is the relaxation time (assumed to be constant), and $f_{\mathrm{eq}}=\exp \left(
-k_{0}/T\right) $ is the local equilibrium distribution function. We use the Landau frame \cite{landau}, the system's energy density is $\varepsilon =3T^{4}/\pi ^{2}$ whereas the pressure $P=\varepsilon/3$. The dynamics of the gas is described by Boltzmann RTA equation above supplemented by the equation for the temperature 
\begin{equation}
\frac{\partial_{\tau}T}{T} +\frac{1}{3\tau}-\frac{\pi}{12\tau}=0,
\label{eqT}
\end{equation} 
where $\pi\equiv \pi^\eta_\eta$ is a component of the shear stress tensor of the fluid defined by the following moment of the distribution function $\pi_\eta^\eta=\int \frac{d^{3}\mathbf{k}}{(2\pi )^{3}\tau }k_0\left[\frac{1}{3}-\left( \frac{k_{\eta }}{k_{0}\tau }\right) ^{2 }\right]f_{%
\mathbf{k}}$. This type of kinetic system has been previously studied in \cite{Baym:1984np} and, more recently, in \cite{Florkowski:2013lza,Florkowski:2013lya}. 

Here we solve (\ref{eqBolt}) and (\ref{eqT}) using the method of moments \cite{Denicol:2012cn}. In this formalism, the underlying kinetic dynamics of the Boltzmann equation is described by an infinite set of moments of $f_\mathbf{k}$ that obey coupled differential equations. Given the symmetries of Bjorken flow \cite{Denicol:2014tha}, we first define the following moments  
\begin{equation}
\rho _{n,\ell }=\int \frac{d^{3}\mathbf{k}}{(2\pi )^{3}\tau }\left(
k^{0}\right) ^{n}\left( \frac{k_{\eta }}{k^{0}\tau }\right) ^{2\ell }f_{%
\mathbf{k}},
\end{equation}%
whose exact evolution equations can be found directly from \eqref{eqBolt}
\begin{equation}
\partial _{\tau }\rho _{n,\ell }+\frac{1+2\ell }{\tau }\rho _{n,\ell }+\frac{%
n-2\ell }{\tau }\rho _{n,\ell +1}=-\frac{1}{\tau _{R}}\left( \rho _{n,\ell
}-\rho _{n,\ell }^{\mathrm{eq}}\right),
\end{equation}%
where 
\begin{equation}
\rho _{n,\ell }^{\mathrm{eq}}=\frac{\left( n+2\right) !}{2\ell +1}\frac{%
T^{n+3}}{2\pi ^{2}}.
\end{equation}%
A nice feature of the method of moments is that one may explore different truncations of the infinite set of moments to systematically improve the description of the fluid \cite{Denicol:2012cn} and directly reconstruct the solution $f_{\mathbf{k}}$ of the kinetic model \cite{Bazow:2015dha,Bazow:2016oky}. To better understand how these moments deviate from their equilibrium values, 
we further define the dimensionless moments 
\begin{equation}
M_{n,\ell }(\tau)\equiv \frac{\rho _{n,\ell }-\rho _{n,\ell }^{\mathrm{eq}}}{\rho
_{n,\ell }^{\mathrm{eq}}}
\end{equation}%
whose equations of motion are
\begin{gather}
\partial _{\tau }M_{n,\ell }+\frac{1}{\tau _{R}}M_{n,\ell }+\frac{6\ell -n}{%
3\tau }M_{n,\ell }-\frac{n+3}{12\tau }M_{1,1}\left(1+ M_{n,\ell}\right)  
+\frac{1}{\tau }\frac{\left( n-2\ell \right) \left( 1+2\ell \right) }{2\ell
+3}M_{n,\ell +1}=-\frac{1}{\tau }\frac{%
4\ell \left( n+3\right) }{3\left( 2\ell +3\right) },
\label{equacaoMnl}
\end{gather}%
where $M_{1,1}=-\pi /P$. This hierarchy of nonlinear equations for the moments may then be numerically solved to determine how the kinetic model evolves in $\tau$. The powerful constraints from Bjorken symmetry, and the simplification made in the collision kernel by employing the RTA, made it possible to determine this general set of equations for arbitrary $n$ and $\ell$. 


\section{Divergence of the Chapman-Enskog expansion}  
\label{CEsection}

\begin{figure}
\centering
\includegraphics[width=0.5\textwidth]{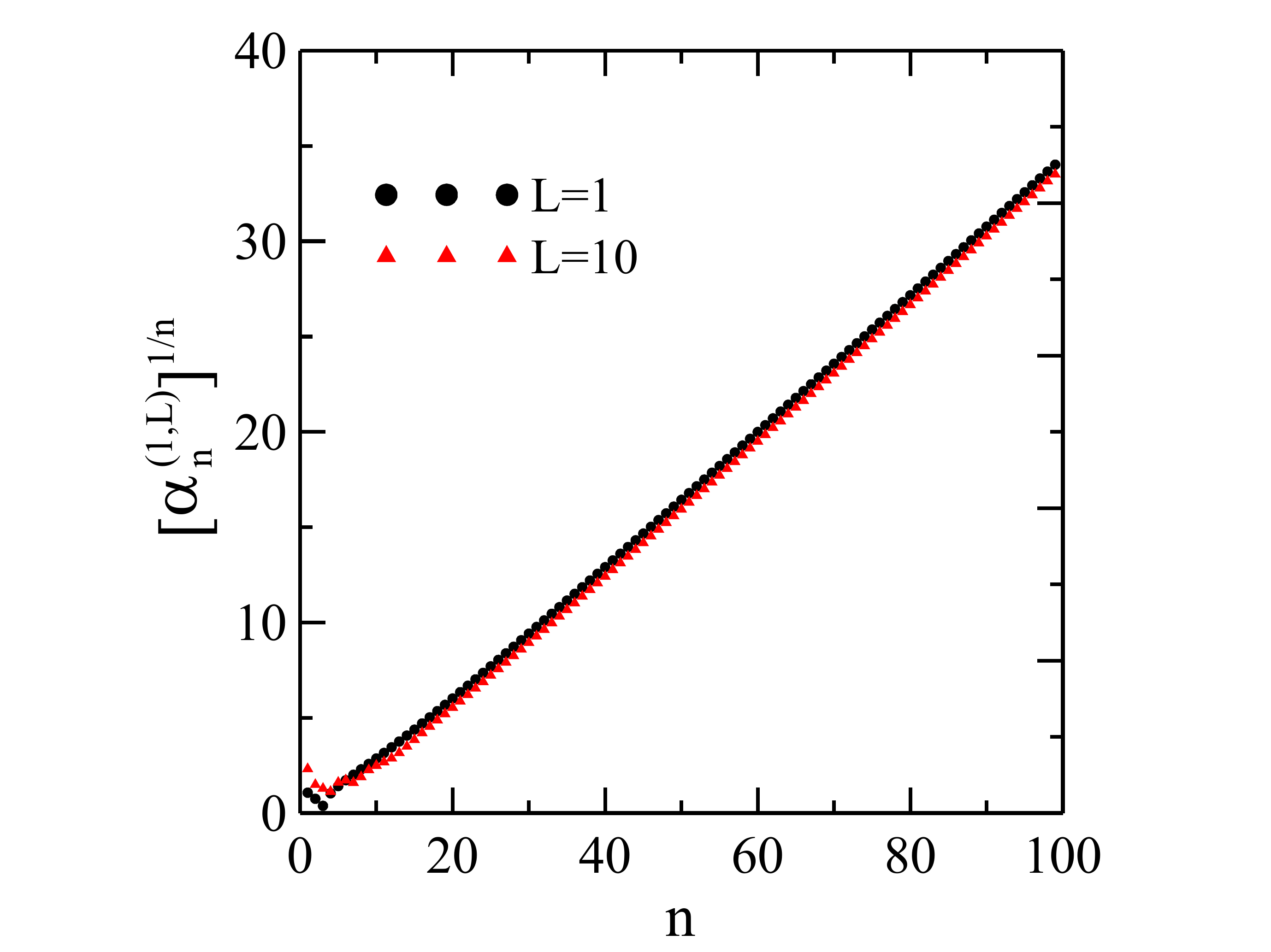}
\caption{(Color online) Coefficients $\left[ \alpha _{m}^{\left( 1,\ell \right) }\right] ^{1/m}$ of the Chapman-Enskog expansion as a function of $m$ for $L=1$ (black circles) and $L=10$ (red triangles).}
\label{divergenceplot}
\end{figure}

The gradient expansion procedure, implemented via the Chapman-Enskog approach, consists in expanding the solution of the Boltzmann equation order by order in powers of the Knudsen number, which for the Bjorken expanding gas considered here corresponds to a series in powers of $K_N \sim \tau_R/\tau$ when $\tau/\tau_R \gg 1$. Given that $f_\mathbf{k}$ may be reconstructed using linear combinations of $M_{n,\ell}$, here we perform the gradient expansion directly in these variables. Therefore, we write    
\begin{equation}
M_{n,\ell }(\hat\tau)=\sum_{p=0}^{\infty }\frac{\alpha _{p}^{\left( n,\ell \right) }}{
\hat{\tau}^{p}},
\end{equation}
where $\hat\tau = \tau/\tau_R$. The series for each moment is characterized by the dimensionless coefficients $\alpha _{p}^{\left( n,\ell \right) }$, which do not depend on $\hat\tau$, and can be determined order by order by the following algebraic equation for $m>1$
\begin{eqnarray}
&&\alpha _{m+1}^{\left( n,\ell \right) }=-\frac{6\ell -n-3m}{3}\alpha
_{m}^{\left( n,\ell \right) }+\frac{n+3}{12}\alpha _{m}^{\left( 1,1\right) } -
\frac{\left( n-2\ell \right) \left( 1+2\ell \right) }{2\ell +3}\alpha
_{m}^{\left( n,\ell +1\right) }+ \frac{n+3}{12}\sum_{p=0}^{m}\alpha
_{p}^{\left( 1,1\right) }\alpha _{m-p}^{\left( n,\ell \right) },
\end{eqnarray}
while $\alpha _{0}^{\left( n,\ell \right) }=0$ and $\alpha _{1}^{\left( n,\ell \right) }=-\frac{4\ell \left( n+3\right) }{
3\left( 2\ell +3\right) }$. A quick look at first term on the right-hand side of the equation above already suggests that $\alpha _{m}^{\left( 1,\ell \right) }\approx
m!$, independently of the value of $\ell$. This can be confirmed by explicit calculation, as depicted in Fig.\ (\ref{divergenceplot}), which shows that the Chapman-Enskog expansion in this case indeed diverges. Other examples where the gradient expansion was found to diverge, though in strongly coupled plasmas that do not admit a kinetic description, can be found in \cite{Heller:2013fn,Buchel:2016cbj}. In hindsight, a simpler argument may be used to show that the gradient expansion considered here diverges. Consider the expansion around $K_N \to 0$. If such an expansion has a nonzero radius of convergence, this means that the series converges in an open region centered at $K_N=0$ in the ``complex" $K_N=\tau_R/\tau$ plane. However, it is clear from \eqref{eqBolt} that the solution is not well defined when $\tau_R \to -\tau_R$ or, in other words, when $K_N \to -K_N$ for any nonzero $\tau_R$. This indicates that the series cannot be well behaved when $K_N \to 0$, having zero radius of convergence (this is similar to Dyson's classical argument for the divergence of perturbative expansions in quantum field theory \cite{Dyson:1952tj}).

In Ref.\ \cite{Denicol:2016bjh} a new type of expansion was proposed to replace the Chapman-Enskog divergent series. This new method involves a generalization of the standard series by simply considering that the coefficients of the expansion may possess a nontrivial $\hat\tau$ (or $K_N$) dependence. More specifically, one assumes   
\begin{equation}
M_{n,\ell }\left( \hat{\tau}\right) =\sum_{p=0}^{\infty } \frac{\beta _{p}^{\left(
n,\ell \right) }\left( \hat{\tau}\right)}{\hat{\tau}^{p}}
\end{equation}%
and determines the \emph{differential equations} obeyed by the coefficients $\beta _{p}^{\left(n,\ell \right) }\left( \hat{\tau}\right)$ extracted order by order in the expansion. While $\beta_{p}^{\left(n,\ell \right) }\left( \hat{\tau}\right)$ asymptotes to $\alpha_p^{(n,\ell)}$ at large $\hat\tau$, one can show that $\beta_{p}^{\left(n,\ell \right) }\left( \hat{\tau}\right)$ obeys nonlinear \emph{relaxation type} equations and, thus, the solutions contain terms such as $ e^{-\hat\tau} \sim e^{-1/K_N}$, which display an essential singularity in $K_N \to 0$. These terms carry the information about the initial condition and they describe the contribution to the dynamics from non-hydrodynamic modes \cite{Kovtun:2005ev}, which are not contained in the original Chapman-Enskog expansion. The generalized equations were numerically solved in \cite{Denicol:2016bjh} and compared to the exact solution of the Boltzmann equation \cite{Florkowski:2013lza,Florkowski:2013lya} and an excellent agreement was found already at the lowest orders in the expansion.

\section{Conclusions}
\label{concl}

In this work we briefly discussed the recent progress towards understanding the emergence of fluid dynamic behavior in rapidly evolving kinetic systems. We focused on the simple example of a Bjorken expanding fluid described by the RTA Boltzmann equation. In this case the Chapman-Enskog series, the well-known expansion defined by powers of the Knudsen number, was shown to diverge. A simple argument involving the behavior of the series in the complex $K_N$ plane was given to explain why such a series has zero radius of convergence. A generalized series, proposed in \cite{Denicol:2016bjh}, reveals that there are non-perturbative contributions to the series, corresponding to non-hydrodynamic degrees of freedom, which cannot be expanded in powers of $K_N$. It would be interesting to consider more realistic collision kernels and investigate the fate of the Chapman-Enskog series and the role played by non-hydrodynamic modes in this case.  


\section*{Acknowledgements}
J.~N. thanks FAPESP and CNPq for support.












\end{document}